
\font\re=cmr8
\font\rn=cmr9

\def\narr{\advance\leftskip by 1.5 pc \advance\rightskip by 1.5 pc}

\hsize=6.1 true in
\vsize=8.7 true in
\nopagenumbers

\topinsert
\vskip 0.3 true in
\endinsert

\def\x{{\bf x}}
\def\sigbar{\bar\sigma}
\def\psibar{\bar\psi}
\def\ve{\varepsilon}

\centerline{\bf BOSONIC PHYSICAL STATES IN $N=1$ SUPERGRAVITY?}
\vskip0.4 true in
\centerline{SEAN M. CARROLL$^a$, DANIEL Z. FREEDMAN$^b$, MIGUEL E. ORTIZ$^c$
AND DON N. PAGE$^d$}
\vskip 0.02 true in
\centerline{\re
$^a$Center for Theoretical Physics and Department of Physics,}
\centerline{\re Massachusetts
Institute of Technology, Cambridge, MA 02139, USA.}
\centerline{\re
$^b$Center for Theoretical Physics and Department of Mathematics,}
\centerline{\re
Massachusetts Institute of Technology, Cambridge, MA 02139, USA.}
\centerline{\re $^c$Institute of Cosmology,
Department of Physics and Astronomy,}
\centerline{\re
Tufts University, Medford, MA 02155, USA.}
\centerline{\re
$^d$CIAR Cosmology Program, Theoretical Physics Institute, Department of
Physics,}
\centerline{\re University of Alberta, Edmonton, Alberta, Canada, T6G 2J1.}
\vskip 0.35 true in
\midinsert
\centerline{ABSTRACT}
\vskip 2pt
\re\baselineskip=10pt
\narr
It is argued that states in $N=1$ supergravity that solve all of the
constraint equations cannot be bosonic
in the sense of being independent of the fermionic degrees of freedom.
(Based on a talk given by Miguel Ortiz at the 7th Marcel Grossmann Meeting.)
\endinsert
\vskip 0.15 true in
\baselineskip 13pt

The canonical quantization of supergravity involves the solution of a
number of constraint equations restricting the form of physical
wave functionals, each of which corresponds to a symmetry of the theory.
The constraint equations for $N=1$ were discussed by D'Eath in 1984$^1$,
where it was shown that in principle it is sufficient to
solve the two supersymmetry constraints in order to obtain completely
gauge invariant wave functionals (this assumes the absence of anomalies
in the operator algebra). The reason for this is that the bracket of the
two supersymmetry constraints yields the familiar Hamiltonian and momentum
constraints that are present because of the diffeomorphism invariance of
the theory. In a subsequent paper which was published this year$^2$,
D'Eath used
this simplifying feature to argue that explicit solutions of the quantum
constraints can be found, and that these have the special property
that they are bosonic. From this result, it has been argued by D'Eath that
supergravity is a finite theory$^3$.
However, this latter claim is dependent upon the
existence of purely bosonic solutions.

In response to Ref.~2, we
demonstrated$^4$ that states solving the
supergravity constraints cannot be independent of the fermionic
variables, and must almost certainly consist of an infinite product of
Grassman valued fields. In this short paper, we review one argument
presented there which shows that a bosonic state of the kind discussed by
D'Eath$^2$
cannot solve the supergravity constraints, and we shall attempt to
update some of the arguments to clarify what is meant in our work by
a bosonic state.

The Lagrangian for $N=1$ supergravity is
$$
{\cal L}={1\over 8\kappa^2}\ve^{\mu\nu\rho\sigma}\ve_{abcd}{E^a}_\mu
{E^b}_\nu {R^{cd}}_{\rho\sigma}-{1\over 2}\ve^{\mu\nu\rho\sigma}
\left(\psibar_\mu {E^a}_\nu\sigbar_a {D_\rho}\psi_\sigma-{D_\rho}
\psibar_\mu {E^a}_\nu\sigbar_a\psi_\sigma
\right)\ ,
\eqno(1)
$$
in terms of Weyl spinor gravitino
fields $\psi_{A\mu}$ and $\bar{\psi}_{A'\mu}$
and a vierbein field
$E^a{}_\mu$. The conventions we use are the same as in Ref.~4.

To work in the canonical formalism,   we must choose a polarisation, and here
we follow Ref.~1 in working in the holomorphic representation, which uses
state functionals depending on $e^a_i$ and $\psi_{Ai}$. As mentioned above,
it is in principle sufficient to look at the supersymmetry constraints,
provided that we ensure that any state functional is a Lorentz invariant
combination of its arguments. These constraints take the relatively simple
form:
$$
\bar{S}^{A'}F=\left[-\ve^{ijk}e^a{}_i{\sigbar_a}^{A'A}
(D_j\psi_{Ak})+{\hbar\kappa^2\over 2}\sigbar^{aA'A}
\psi_{Ai}{\delta\over\delta e^a{}_i}\right]F[{e^a}_i,\psi_{Ai}]=0\
\eqno(2)
$$
and
$$
S^AF=\left[D_j{\delta\over\delta\psi_{Aj}}+{\hbar\kappa^2\over
2}{\delta\over\delta\psi_{Bj}}D_{BA'ji}\sigbar^{aA'A}{\delta\over\delta
e^a{}_i}\right]F[{e^a}_i,\psi_{Ai}]=0\ ,
\eqno(3)
$$
where the $\sigma$ and $\bar\sigma$ are sigma matrices and $D_{AA'ij}$
is a function of ${e^a}_i$. The connection for
the covariant derivative $D_j$ and other details can be found in Ref.~4.

The important feature of the constraints (2) and (3) is that every term in
(3) involves derivatives with respect to the gravitino field, so any state
that is independent of $\psi_{Ai}$,
$$
{\delta F[{e^a}_i,\psi_{Ai}]\over\delta\psi_{Ai}}=0\ ,
\eqno(4)
$$
automatically solves (3). Thus if a solution of the form $F[{e^a}_i,
\psi_{Ai}]=F^{(0)}[{e^a}_i]$
can be found to Eq.~(2), it is a physical state.  We now present a simple
scaling argument from Ref.~4 which shows that there is no such solution
of (2).

\def\x{{\bf x}}

Multiplying by $F^{-1}$ and integrating over
an arbitrary continuous spinor test function
$\bar{\epsilon}(\x)$, the constraint (2) becomes
$$
\int d^3\x\ \bar{\epsilon}(\x)\left[
-\ve^{i j k}{e^a}_i(\x)\sigbar_a (D_j\psi_k(\x))+{\hbar\kappa^2\over
2}\psi_i(\x)\sigbar^a{\delta(\ln F^{(0)}[e])
\over\delta{e^a}_i(\x)}\right]=0\ ,
\eqno(5)
$$
which must be satisfied
for all $\bar{\epsilon}(\x)$, ${e^a}_i(\x)$, and $\psi_{k}(\x)$.
Let the integral in Eq.~(5) be $I$, and let $I'=I+\Delta I$ be the
integral when $\bar{\epsilon}(\x)$ is replaced by
$\bar{\epsilon}(\x)e^{-\phi(\x)}$ and $\psi_{i}(\x)$ is replaced by
$\psi_{i}(\x)e^{\phi(\x)}$, where $\phi(\x)$ is a scalar function.
Since $\bar{\epsilon}(\x)\psi_{i}(\x)$
is unchanged, the second term (with the
functional derivative) cancels in the difference between $I'$ and $I$,
so that we must have
$$
\Delta I=-\int d^3\x\
\ve^{ijk}{e^a}_i(\x)\bar{\epsilon}(\x)\sigbar_a\psi_k(\x)
\partial_j\phi(\x)=0\ .
\eqno(6)
$$
Notice that $\Delta I$ is independent of the state $F^{(0)}[e]$.
Clearly, it is possible to choose the arbitrary fields
$\bar\epsilon(\x)$, $\phi(\x)$, $e^a{}_i(\x)$ and $\psi_k(\x)$ such that
(6) is nonvanishing; therefore no physical state is bosonic in the sense
of Eq.~(4).

The constraint equation (2) can in principle be solved using the method
of characteristics$^2$. In this approach, the value
of the wave functional is specified
on an appropriate subspace of the configuration
space $\{ {e^a}_i,\psi_{Ai}\}$ and the constraint (2) is used to calculate
the value throughout the rest of the space.  By the scaling argument
above, such a construction can never yield a state which is independent
of $\psi_{Ai}$ throughout configuration space;
it would thus be necessary to check independently
that the state was a solution of the $S^A$ constraint (3), which was not
carried out in Ref.~2.

We now show directly why the method of characteristics fails to produce
bosonic solutions.  Assume that at some fixed configuration
${e^a}_i{}^{(0)}(\x)$ the
functional $F[e^{(0)},\psi_{Ai}]$
is independent of $\psi_{Ai}$.  An infinitesimal
$\bar S$ supersymmetry transformation changes ${e^a}_i{}^{(0)}(\x)$ to
$$
{e^a}_i{}^{(0)}+\delta {e^a}_i \equiv
{e^a}_i{}^{(0)}-{i\kappa^2\over 2}{\bar\epsilon}_{A'}\bar\sigma^{aA'A}
\psi_{Ai}\ ,
\eqno(7)
$$
leaving $\psi_{Ai}$ unchanged.  From (2) we see that the value of $F$
at the transformed configuration is
$$
F[e^{(0)}+\delta e,\psi] = F[e^{(0)}]
-\int d^3\x\ \bar\epsilon_{A'}\varepsilon^{ijk}{e^a}_i
{{\bar\sigma}_a}^{A'A}D_j\psi_{Ak} F\ .
\eqno(8)
$$
Scaling $\psi$ and $\bar\epsilon$ as before leaves $\delta e$ unchanged
and thus gives
$$
F[e^{(0)}+\delta e,e^{\phi(\x)}\psi] = F[e^{(0)}]
-\int  d^3\x\ e^{-\phi(\x)}\bar\epsilon_{A'}\varepsilon^{ijk}
{e^a}_i{{\bar\sigma}_a}^{A'A}D_j(e^{\phi(\x)}\psi_{Ak}) F\ .
\eqno(9)
$$
The right hand sides of (8) and (9) will not be equal in general,
which shows that the state $F$ cannot be independent of $\psi$ in
the full configuration space of the theory.

We conclude that any solution
of the constraint (2) arrived at using the method of characteristics cannot
be bosonic in the sense we have defined. Therefore only a very small subclass
of solutions of this constraint can also be solutions of (3);
Ref.~4 presents arguments that any true physical state must
contain an infinite product of Grassmann fields.

\vskip 0.4 true in
\centerline{\bf Acknowledgements}
\vskip 0.1 true in
This short paper draws heavily on material contained in Ref.~4
and should be regarded as a summary of certain aspects of that work.
We thank Peter D'Eath and Hermann Nicolai for helpful discussions.

\vskip 0.4 true in

\noindent{\bf References}
\vskip 0.1 true in
{\rn
\item{1.} P. D. D'Eath, {\sl Phys. Rev.}  {\bf D29} (1984) 2199.

\item{2.} P. D. D'Eath, {\sl Phys. Lett.} {\bf B321} (1994) 368.

\item{3.} P. D. D'Eath, private communication.

\item{4.} S. M. Carroll, D. Z. Freedman, M. E. Ortiz and D. N. Page,
{\sl Nucl. Phys.} {\bf B423} (1994) 661.
}
\end